# Wideband, Efficient AlScN-Si Acousto-Optic Modulator in a Commercially Available Silicon Photonics Process

Mertcan Erdil[1,2], Izhar[1,2], Yang Deng[1], Zichen Tang[1], Mohamad Hossein Idjadi[1], Farshid Ashtiani[1], Firooz Aflatouni[1,3] & Roy Olsson III[1,3]

Acousto-optic integration offers numerous applications including low-loss microwave signal processing, nonreciprocal light propagation, frequency comb generation, and broadband acousto-optic modulation. State-of-the-art acousto-optic systems are mainly implemented entirely using in-house fabrication processes, which despite excellent performance typically suffer from low yield and are not compatible with mass production through foundry processes. Here, we demonstrate a highly efficient wideband acousto-optic modulator (AOM) implemented on a silicon photonics foundry process enabling high-yield low-cost mass production of AOMs with other photonic and electronic devices on the same substrate. In the reported structure, a 150 μm long AlScN-based acoustic transducer launches surface acoustic waves (SAW), which modulate the light passing through a silicon optical waveguide. A modulation efficiency of -18.3 dB over a bandwidth of 112 MHz is achieved, which to our knowledge is the highest reported efficiency and bandwidth combination among silicon based AOMs, resulting in about an order of magnitude $BW(V_\pi L)^{-1}$ figure-of-merit improvement compared to the state-of-the-art CMOS compatible AOMs. The monolithically integrated acousto-optic platform developed in this work will pave the way for low-cost, miniature microwave filters, true time delays, frequency combs, and other signal processors with the advanced functionality offered by foundry-integrated photonic circuits.

Acousto-optic systems harnessing phonon-photon interactions offer significant advancements in many applications such as microwave photonics, optical frequency shifting, non-reciprocal photonics, sensing, spectroscopy, and quantum information control[1–4]. Combining mechanical wave technology with integrated photonic systems enables the implementation of advanced devices for radio-frequency (RF) and optical signal processing[3,5,6]. An important example is the acousto-optic modulator (AOM), which is a key component for the transduction of information between the acoustic and optical domains.

Conventionally, acousto-optic modulators are implemented using bulk-crystalline materials, which despite excellent performance, are often bulky, expensive and not suitable for large scale integrated applications due to a rather weak acoustic wave confinement[2,6,7].

Surface acoustic wave (SAW) based modulators offer a high degree of acoustic wave confinement on the surface of the structure and can be co-integrated with photonic devices on the same platform[5,7–9]. The operation principle of a SAW modulator relies on the photo-elastic effect[10], where the mechanical wave launched by an interdigitated transducer (IDT), implemented using a piezoelectric layer, interacts with an optical waveguide and modulates the refractive index of the waveguide medium as a result of the applied strain[11]. Due to a high degree of confinement and a large overlap between the acoustic and optical fields, efficient SAW based AOMs in a small footprint can be realized[7,8,12,13]. Such modulators have been implemented using different material systems. Lithium niobate platforms offer strong piezoelectric and electro-optic properties and have been used for implementation of SAW based AOMs[6,8,9,14–17]. Despite outstanding efficiency, lattice mismatch between lithium niobate and silicon makes the large-scale integration of such devices with silicon photonics and complementary metal-oxide semiconductor (CMOS) electronics challenging[12]. Similarly, PZT based AOMs offer a high acousto-optic modulation efficiency[11] but are not compatible with CMOS platforms.

[1]Department of Electrical and Systems Engineering, University of Pennsylvania, PA, USA. [2]These authors contributed equally: Mertcan Erdil, Izhar. [3]Corresponding authors: e-mail: firooz@seas.upenn.edu, e-mail: rolsson@seas.upenn.edu

Aluminum nitride (AlN) is another piezoelectric material that has been used for integration of SAW transducers on silicon photonic chips[7,13,18,19], where despite compatibility with CMOS processes, the relatively weak piezoelectric response of AlN limits the efficiency of the AlN AOMs. The piezoelectric response of AlN can be improved through scandium alloying (resulting in AlScN) while maintaining its compatibility with CMOS and silicon photonics integration[12,20–22].

To date, AlScN based SAW transducers integrated with photonic systems are entirely based on in-house photonic fabrication processes, which despite innovative approaches, suffer from limited yield, high cost, and incompatibility with mass production.

Here we demonstrate the first AlScN-based acoustic transducer implemented using a commercial silicon photonics process. To our knowledge, compared to other reported CMOS compatible AOMs, this work achieves the highest measured acousto-optic modulation efficiency, and does so over more than five times larger operating bandwidth. The AOM device is implemented by post processing directly on photonic structures fabricated using the AMF silicon-on-insulator (SOI) photonics process.

The implemented modulator achieves a modulation efficiency of -18.3 dB measured at 23.2 dBm RF input power, corresponding to a $V_\pi L$ of 1.12 V-cm. Furthermore, the device exhibits a modulation bandwidth of 112 MHz (around 5.5 GHz) resulting in about an order-of-magnitude $BW(V_\pi L)^{-1}$ figure-of-merit improvement over state-of-the-art reported silicon-based CMOS compatible AOMs.

The demonstrated integration of highly efficient acoustic devices on a CMOS-compatible commercial photonics process enables mass production of high-yield, low-cost acoustic, photonic, and CMOS devices on the same chip, resulting in advances in many applications from sensing and imaging to signal processing and computation.

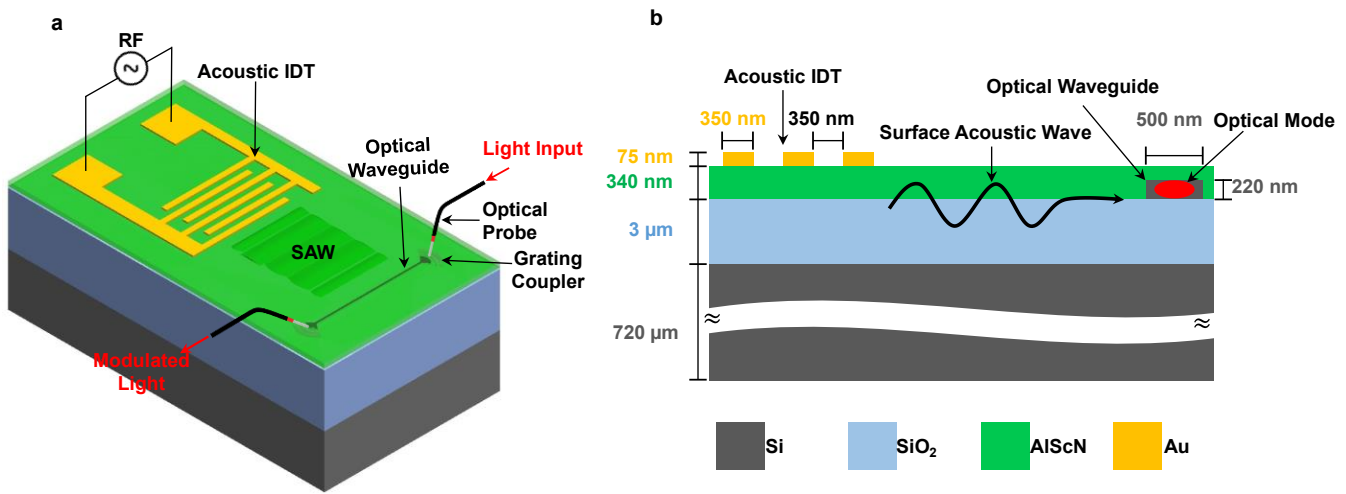

**Fig. 1 | Principle of operation of the acousto-optic modulator. a,** structure and **b,** cross-sectional view of the device. The surface acoustic wave (SAW) launched by an acoustic interdigitated transducer propagates and produces strain when impinging upon the optical waveguide. The strain induced refractive index change results in phase modulation of the optical wave. IDT: interdigitated transducer.

## Results

The schematic and cross-section of the implemented acousto-optic modulator are shown in Figs. 1a and 1b, respectively. The AOM structure consists of a single-mode optical waveguide fabricated in AMF SOI process and an acoustic transducer fabricated next to the optical waveguide using a CMOS compatible, in-house post processing step. The acoustic transducer is implemented by first depositing an AlScN piezoelectric film on the buried oxide layer (BOX) of the SOI chip followed by fabrication of gold interdigitated finger electrodes over the AlScN layer. When an RF signal is applied to the acoustic transducer, a SAW is launched that propagates on the surface of the BOX and interacts with the optical waveguide. As a result, the optical waveguide experiences strain which causes a change in the refractive index of the waveguide proportional to the acoustic wave strength. This acoustic induced refractive index change modulates the phase of the optical wave travelling within the waveguide. As a result of this, acousto-optic phase modulation sidebands appear around the optical carrier. The ratio of the power in the first upper or lower sideband to the power in the carrier is defined as the AOM efficiency ($\eta_{AOM}$) and is written as

$$\eta_{AOM} = 20\log\left(\frac{\alpha(L)}{2}\right) = P(\omega_o \mp \Omega)_{dB} - P(\omega_o)_{dB}, \quad (1)$$

where $\omega_o$, $\Omega = 2\pi f$, $f$, $\alpha(L) = -2\pi \Delta n_{eff} L \lambda_0^{-1}$, $\lambda_0$, $L$ and $\Delta n_{eff}$ are the angular frequency of the optical wave, the angular frequency of the RF signal, the frequency of the RF signal, the amount of optical phase shift due to acousto-optic modulation, the wavelength of the optical wave, the interaction length and the acoustically induced effective index change, respectively.

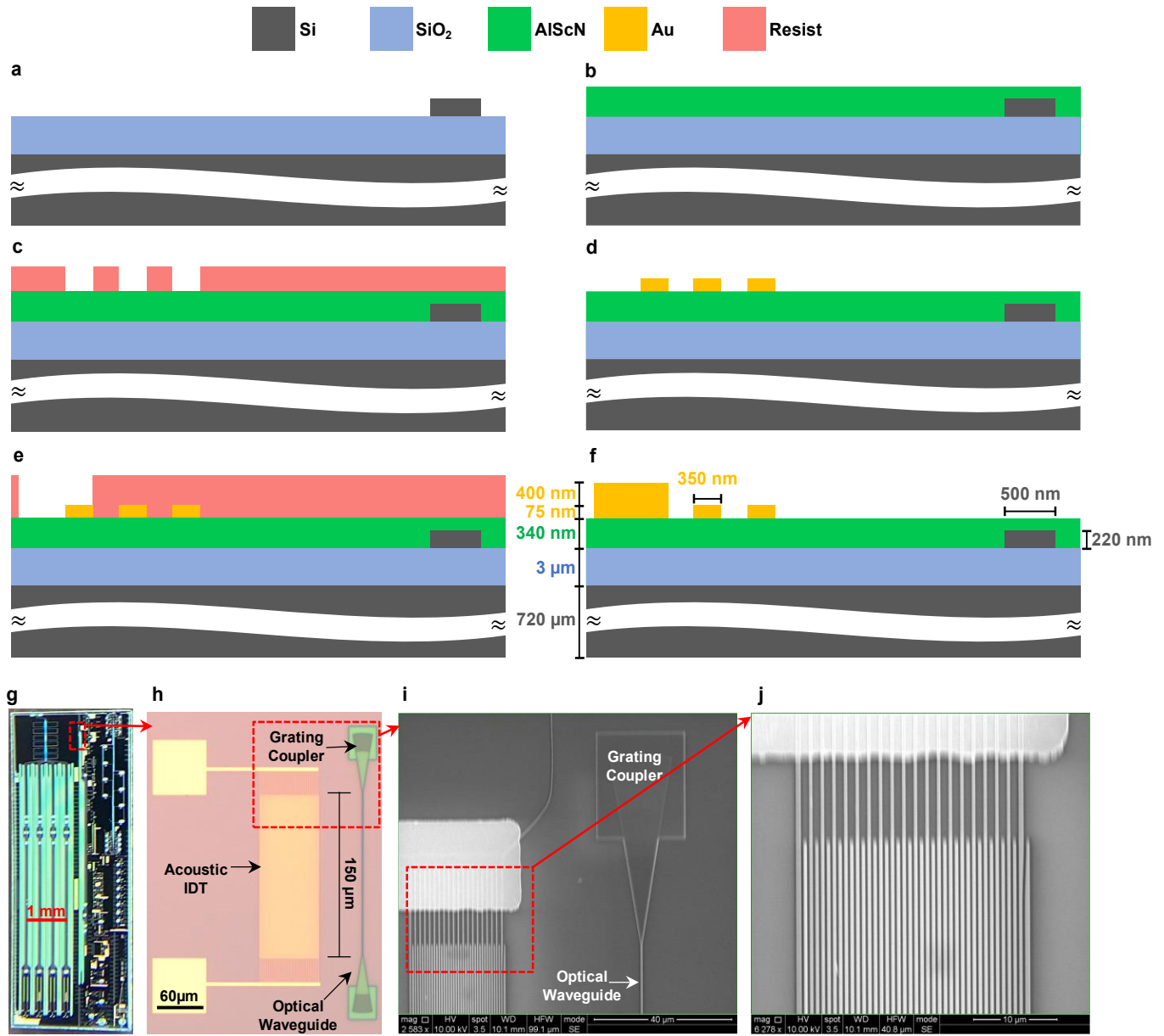

**Fig. 2 | Fabrication of the acoustic device on the photonic chip. a,** Cross-section of the chip fabricated with oxide opening in AMF foundry process. **b,** Sputtering of AlScN on the chip. **c,** Deposition and patterning of ebeam resist. **d,** Evaporation and patterning of Au interdigitated fingers using lift-off process. **e,** Spin coating and development of photoresist for Au bus lines and pads. **f,** Patterning of Au for bus lines and pads. **g,** Micrograph of the chip fabricated using AMF SOI platform. **h,** Acoustic transducer fabricated on the chip near optical waveguide for AOM. **i,** Scanning electron microscope (SEM) image of acoustic transducer and optical waveguide. **j,** Zoomed-in view of Au IDT.

The cross-section of the photonic chip fabricated in the AMF process is shown in Fig. 2a. To implement the acousto-optic modulator, first, a 340 nm thick layer of AlScN is sputtered on the exposed Si on SiO$_2$ layer of the photonic chip (Fig. 2b) using a pulsed DC physical vapor deposition (PVD) system[23–26]. The AlScN deposition was followed by spin coating and patterning of e-beam resist on the chip to pattern the interdigitated electrodes (Fig. 2c). Next, a 75 nm thick layer of Au (with 10 nm titanium (Ti) as adhesion layer) is evaporated and patterned to fabricate the IDT electrodes using the lift-off process (Fig. 2d). Note that the 700 nm pitch of the IDTs is optimized to excite a SAW at 5.5 GHz. Photoresist was spin coated and patterned to form low resistance bus lines and pads for the acoustic transducer. Finally, a 400 nm thick layer of Au (with 10 nm Ti as adhesion layer) was evaporated and patterned using lift-off to fabricate the bus line and pads for the acoustic transducer.

A micrograph of the fabricated acousto-optic device and its zoomed-in version are shown in Figs. 2g and 2h, respectively. The SEM of the fabricated device and its zoomed-in version are shown in Figs. 2i and 2j, respectively. The analysis on the quality of the AlScN deposition is presented in Supplementary Fig. 1.

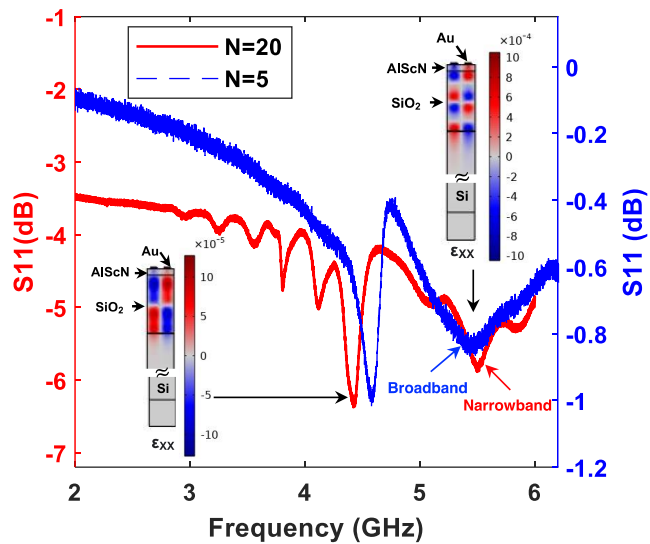

**Fig. 3| IDT Characterization.** The electrical reflection coefficient (S$_{11}$) response of the acoustic transducers fabricated on the commercial photonic chip and FEM simulations showing the strain at resonant frequencies around 4.5 GHz and 5.5 GHz (for both transducers). Simulations confirm that both modes launch SAWs as the strain fields are confined in the top layer stack (Au/AlScN/Oxide) on the chip. Due to a larger strain, the 5.5 GHz mode is expected to have higher AOM efficiency. The response of the transducer with 5 finger pairs has a larger bandwidth, however, the device with 20 finger pairs is better matched to the source impedance.

The fabricated acoustic transducer on the commercial photonic chip was characterized using a vector network analyzer, VNA (Keysight E8361A), which was operated at an input power of -10 dBm and a port impedance of 50 $\Omega$. A one-port frequency sweep from 2 GHz to 6 GHz using a ground-signal (GS) probe was performed. The VNA was calibrated to the probe tips using the open-short-load procedure prior to testing of the acoustic transducer. The electrical reflection coefficient (S$_{11}$) response of the acoustic transducer as a function of frequency is shown in Fig. 3, where the transduction of the RF power into acoustic waves appear as dips (corresponding to the frequency of the excited SAW) in the reflected power spectra.

The transducers, as illustrated in Fig. 3, show two rather strong responses at frequencies of approximately 4.5 GHz and 5.5 GHz, which follows the $sin(x)/x$ profile as expected[27–29]. To study the effect of number of IDT finger pairs (N) on the acousto-optic modulation bandwidth, two SAW transducers with different number of IDT finger pairs were fabricated. As evident from the $S_{11}$ measurements in Fig. 3, while the IDT with 20 finger pairs is better matched to the source impedance compared to the transducer with 5 finger pairs, the latter has a larger bandwidth, which is expected as the bandwidth of the SAW transducer is inversely proportional to the number of IDT finger pairs[28–30].

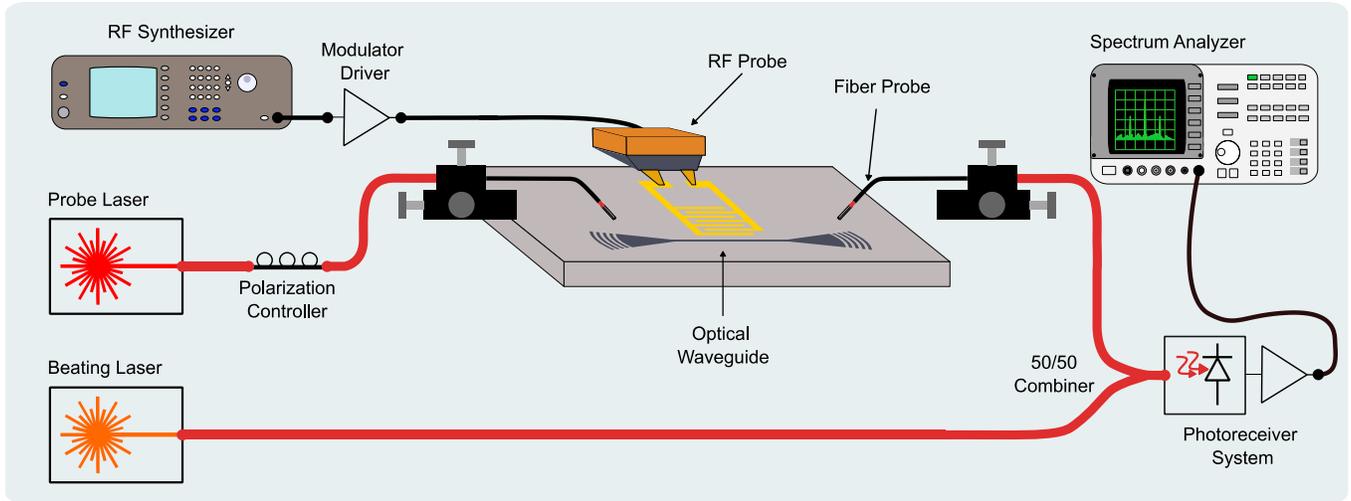

**Fig. 4 | Experimental Setup.** The heterodyne setup built for AOM efficiency measurements. The probe laser is coupled into the chip via a fiber probe and a grating coupler. The output of an RF synthesizer is amplified and used to drive the IDT. The optical output of the chip is combined with the output of a heterodyning laser (operating at a slightly different wavelength) via a 50/50 fiber combiner. The combiner output is photo-detected and amplified by a photoreceiver system. Finally, the modulated signal is analyzed using a spectrum analyzer. In the schematic, the red and black wires indicate the optical and electrical paths, respectively.

The propagation modes for the two IDTs were simulated using COMSOL Multiphysics (Fig. 3). The simulation results confirm the presence of two dominant modes at 4.5 GHz and 5.5 GHz. Note that while both modes support propagation of surface acoustic waves (as the strain fields are confined in the Au/AlScN/Oxide layer stack on the chip), the strain in the silicon waveguide associated with the mode at 5.5 GHz is stronger than that of the mode at 4.5 GHz. This is due to the fact that given the velocity of sound in the medium, $v$, IDT pitch, $p = \frac{\lambda}{2} = \frac{v}{2f}$, was designed to excite the mode at $f = 5.5$ GHz, where $\lambda$ is the acoustic wavelength[29,31]. The velocity of the SAW was first calculated using COMSOL simulations and then it was used to adjust the pitch of the IDTs to optimize the transducer response at 5.5 GHz. Since the propagating SAW at 5.5 GHz mode has a larger strain in this case, a relatively stronger acousto-optic modulation is expected at this frequency.

The modulation efficiency of the implemented AOM device was measured using a heterodyne detection scheme as depicted in Fig. 4.

The output of a tunable laser (Agilent 8164A), emitting 5 dBm at 1550 nm, is polarization adjusted and coupled into the chip using a grating coupler. The output of an RF synthesizer is also coupled into the chip after being amplified by a modulator driver and is used to acousto-optically modulate the coupled light. The modulated light is coupled out of the chip, combined with the output of another tunable laser (Agilent 8164A) and photo-detected and amplified by a photoreceiver system (Thorlabs RXM40AF). The difference between the wavelength of the two laser sources was set such that their beat note frequency is around 7 GHz. To find the frequency response of the AOM, the frequency of the RF synthesizer was swept from 2 GHz to 6 GHz. The heterodyne spectrum for the device with 20 fingers, normalized to the beat-note power, is shown in Fig. 5a (for an RF power of 23.2 dBm), where the main resonant peak appears around 5.5 GHz. The modulation efficiencies of the first lower and upper sidebands around the resonant peak are shown in Fig. 5b, where the modulation response is symmetric as predicted by equation (1). The slight difference between the upper and lower sideband modulation responses is attributed to the frequency dependent loss of the

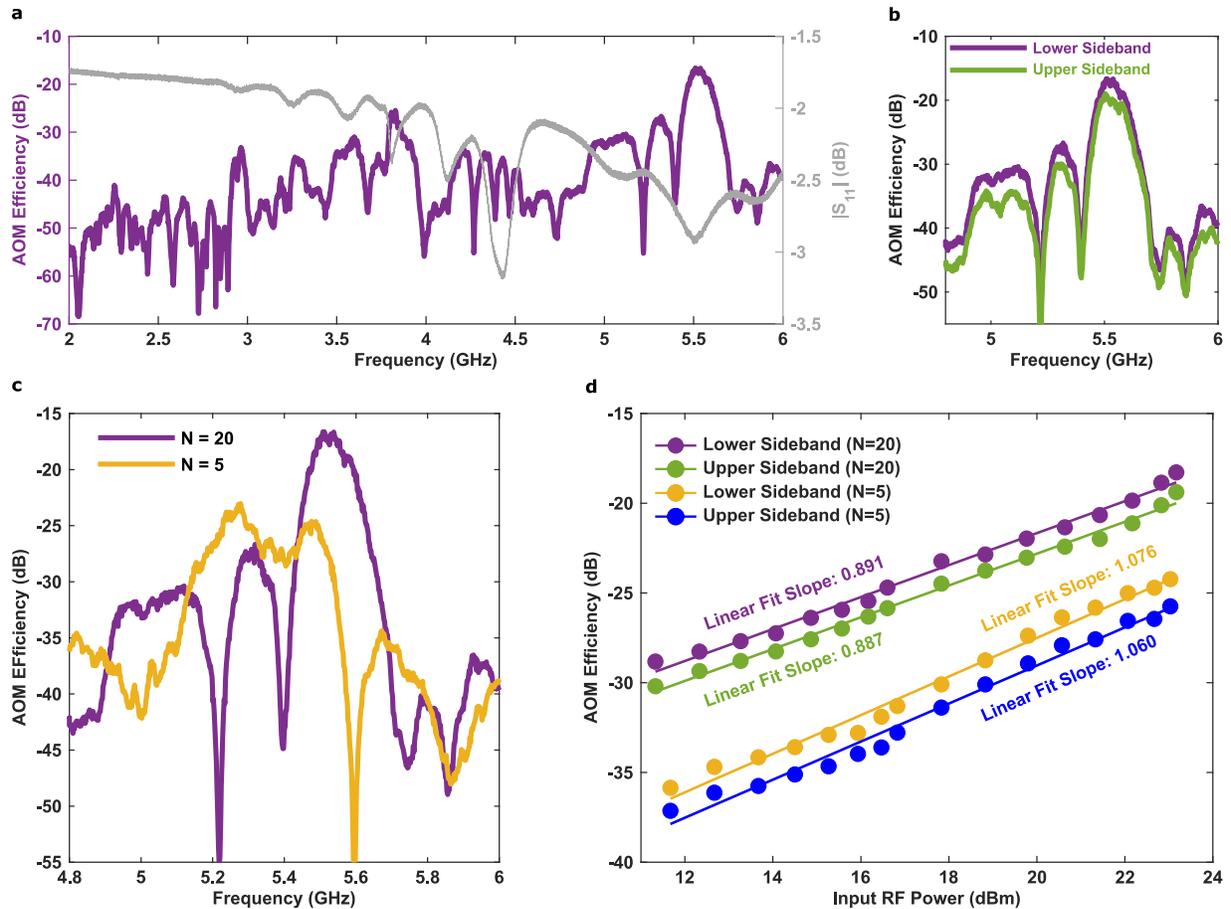

**Fig. 5 | AOM efficiency measurements. a,** The frequency spectrum of the AOM efficiency (output power normalized to beat-note power) of the modulator together with the $S_{11}$ response of the IDT device with 20 finger pairs. **b,** Acousto-optic modulation efficiency of lower and upper sideband signals around the resonance peak for the device with 20 fingers. **c,** The comparison of AOM response of the devices with number of finger pairs of 20 and 5. The purple and yellow lines presents the zoomed-in AOM efficiency for the IDT device with 20 and 5 finger pairs, respectively. **d.** The AOM efficiency characteristics of the lower and upper sidebands versus the RF signal power. Purple (yellow) and green (blue) circles represent the AOM efficiency measurements for the lower and upper sidebands for IDT device with 20 and 5 finger pairs, respectively, while purple (yellow) and green (blue) lines correspond to the linear fit to the data. Slope of ~1 corresponds to a linear relationship between the measured efficiency and the applied input RF power.

---

experimental setup, since the two sidebands differ in frequency by an amount of $2f$, where $f$ is the acoustic wave frequency.

To study the effect of the number of IDT finger pairs on device performance, we have implemented other IDT devices with 5 finger pairs and compared the performance of the IDT devices with 20 finger pairs and 5 finger pairs. The results are shown in Fig. 5c, where, as expected, a lower number of IDT finger pairs results in a higher operation bandwidth at the cost of a lower AOM peak efficiency. Here, after de-embedding the cable loss, at 23.2 dBm RF power, the IDT device with 20 finger pairs achieves a peak AOM efficiency of -18.3 dB with a bandwidth of 112 MHz, while the device with 5 finger pairs achieves a peak AOM efficiency of -24.2 dB with a bandwidth of 294 MHz, confirming the expected trade-off between the peak AOM efficiency and bandwidth for different number of IDT finger pairs.

The effect of RF power on the acousto-optic modulation efficiency was also studied, where the RF power was swept and the AOM efficiency for the lower and upper sidebands were measured (Fig. 5d). As suggested by theory[11], the power of both lower and the upper modulation sidebands increases linearly with RF power. For the device with 20 finger pairs (N=20), the lower sideband modulation efficiency of -18.3 dB and an upper sideband modulation efficiency of -19.4 dB at an RF power of 23.2 dBm were achieved, respectively. Based on the efficiency measurements, a $V_\pi L$ of about 1.12 Vcm for this modulator is calculated. For the device with 5 finger pairs (N=5), the modulation efficiency values of -24.2 dB and -25.7 dB were measured for lower and upper sidebands, respectively. Corresponding $V_\pi L$ is calculated to be around 3.25 Vcm. Note that the modulator is still operating in the linear regime at 23.2 dBm input RF power (Fig. 5d), indicating that, if needed for an application, even higher modulation efficiencies could be achieved at higher RF powers.

## Discussion

The performance of the AOM devices developed in this work is compared with the performance of other works in Table 1. The IDT with 20 finger pairs achieves a modulation efficiency of about -18.3 dB, which is the highest among all devices reported in Table 1. The BW exhibited by our 20-finger pair device (112 MHz) is more than 5 times higher when compared to other devices realized in CMOS-compatible processes. On the other hand, our 5-finger pair device exhibits more than an order of magnitude improvement on the operation bandwidth compared to the other reported works. Additionally, our devices can operate at a higher frequency (5.5 GHz) compared to other reported works (0.576 to 3.11 GHz). Note that our integrated AOM platform is the only one which utilizes a CMOS-compatible fabrication process to integrate acoustic transducers on foundry fabricated chips. An important figure-of-merit for electro-optic modulators is $V_\pi L$, which captures both the modulation efficiency, power consumption and the footprint of the device. However, it does not capture bandwidth, which is an important metric for often narrow band AOMs. There exists a clear trade-off between the modulation efficiency and the bandwidth in SAW-based acousto-optic modulators, as our results here confirmed. Therefore, a figure-of-merit that is better suited for AOMs can be defined as $FOM II = BW/V_\pi L$, which combines the modulation efficiency, power consumption, device footprint and the bandwidth together. As shown in Table 1, our device with 20 finger pairs has about an order-of-magnitude higher FOM II compared to other works.

The performance of the implemented transducer could be further improved by fabricating an acoustic reflector to realize an unidirectional transducer with a higher efficiency[32]. Additionally, a matching network can be added to the input of the transducer

Table 1| Table of comparison for silicon based AOM platforms.

| Ref. | Material | Platform | Post Processing | $\eta_{AOM}$ (dB) | BW (MHz) | N | f (GHz) | $V_\pi L$ (V.cm) | $\frac{BW}{V_\pi L}\left(\frac{MHz}{V.cm}\right)$ |
|---|---|---|---|---|---|---|---|---|---|
| Ansari et al.[11] (2022) | PZT | Foundry Fabricated Si-Photonics | Non-CMOS Compatible | -36.5 (L =180 µm) | NR | 20 | 0.576 | 3.35 | NR |
| Ansari et al.[11] (2022) | PZT | Foundry Fabricated Si-Photonics | Non-CMOS Compatible | -43.8 (L =70 µm) | NR | 4 | 2 | 3.6 | NR |
| Huang et al.[12] (2022) | AlScN | In-house SOI | CMOS Compatible | -20.6 (L =210 µm) | 11 | 100 | 3.375 | 0.95 | 11.58 |
| Huang et al.[12] (2022) | AlScN | In-house SOI | CMOS Compatible | -21.5 (L =210 µm) | 4 | 100 | 3.044 | 1.06 | 3.77 |
| Kittlaus et al.[13] (2021) | AlN | In-house SOI | CMOS Compatible | -20 (L =240 µm) | 20 | 107 | 3.11 | 1.8 | 11.11 |
| **This Work (N=20)** | **AlScN** | **Foundry Fabricated Si-Photonics** | **CMOS Compatible** | **-18.3 (L =150 µm)** | **112** | **20** | **5.5** | **1.12** | **100** |
| **This Work (N=5)** | **AlScN** | **Foundry Fabricated Si-Photonics** | **CMOS Compatible** | **-24.2 (L =150 µm)** | **294** | **5** | **5.3** | **3.25** | **90.46** |

*NR*: Not Reported, N: number of fingers.

to improve the RF conversion efficiency[12]. The developed AOM platform implemented on a commercial Si-photonics process can pave the way for the development of high-performance, high-yield CMOS-compatible AOMs bringing advanced acoustic signal processing capabilities to Si-CMOS based photonic and electronic platforms.

## Methods

### Fabrication of photonic chip

The photonic devices were fabricated using AMF SOI process with a 3 µm thick buried oxide layer. The transmission loss of the photonic structure is mainly dominated by the grating coupler coupling loss (~5 dB). No significant change in the optical loss was observed after AlScN deposition.

### Fabrication of acoustic IDTs

Acoustic IDTs were produced on the photonic chip using in-house microfabrication process. The chips received from AMF foundry were first cleaned using acetone, isopropyl alcohol (IPA) and deionized water followed by nitrogen drying prior to growth of AlScN piezoelectric layer. During the AlScN growth, a shadow mask was used to protect certain areas of the photonic chip where AlScN growth was not desired. The targets in PVD system were conditioned to ensure a high quality AlScN deposition. In the PVD system, separate 4-inch diameter Al and Sc targets and nitrogen gas were used to deposit the AlScN. Before the deposition, the process chamber was pumped to a base pressure of approximately $9 \times 10^{-8}$ mbar and the substrate temperature was raised to 350 °C. To achieve the desired composition of 30% Sc in the deposited film, during the deposition, respective cathode powers of 555 W and 1000 W for Sc and Al targets were used while the nitrogen flow was set to 20 sccm.

The ebeam resist used for IDTs patterning was ZEP 520A, which was spin coated on the chip at 2200 rpm and backed at 120 °C temperature for 5 minutes before exposer in ebeam lithography. Microposit remover 1165 at 65 °C temperature was used for liftoff to pattern Au IDT fingers.

To fabricate pads and bus lines, microposit S1800 series photoresist was spin coated at 3000 rpm, backed at 125 °C, expose and developed in microposit MF CD-26 developer at room temperature. The same microposit remover 1165 at 65 °C was used for lift-off.

### Characterization of AlScN Deposition

To evaluate the quality of the AlScN deposited on the commercial photonic chip, atomic force microscope (AFM) analysis of the deposited AlScN was performed as shown in Extended Data Fig. 1a. Overall, the surface roughness of the film was approximately 1.4 nm and very few abnormally oriented grains (AOGs) were observed. Furthermore, the X-ray diffraction (XRD) omega-scan rocking curve of the deposited film on the photonic chip was measured to evaluate the film quality as shown in Extended Data Fig.1b. The full width half maximum (FWHM) of 2.92° was extracted through peak fitting. Note that although most of the film in this work was grown on the amorphous $SiO_2$ layer (which dominates the exposed surface of the Si photonics chip), the measured FWHM is similar to those reported for similar AlScN thicknesses grown on Si[23,33] (*i.e.* 2.1-2.4°). Energy dispersive spectroscopy (EDS) analysis of the AlScN thin film confirmed 30% Sc in the AlScN alloy.

### Characterization of acoustic IDTs

To characterize the acoustic IDTs, a Keysight E8361A VNA and a ground-signal (GS) probe were used in the setup. The VNA was calibrated at the probe tips using the standard open-short-load (OSL) technique.

### Heterodyne AOM measurements

The heterodyne measurements are performed by beating two lasers with different frequencies. The wavelength of the probe laser passing through the chip was set to 1550 nm, whereas the heterodyning laser wavelength was set to 1550.058 nm, resulting in a beat-note frequency of around 7 GHz. When two beams are combined and photo-detected, two sidebands around the 7 GHz beat-note are observed on the spectrum analyzer. The sidebands and beat-note levels are measured for AOM efficiency calculations. For each RF power level, multiple data points are recorded, and the average value of the efficiency is reported. The efficiency values in Fig. 5d are reported after de-embedding the frequency dependent cable transmission loss.

## Calculation of V$_\pi$L product

To calculate $V_\pi$ of the modulators, the impedance of the devices is extracted by utilizing the measured S$_{11}$ data. Then, the voltage across the devices ($V_{dev}$) is calculated when they are driven by an RF source with 50 Ω output resistance. The voltage level of the source is determined by utilizing the power level ($P_{RF}$) measured when the output of the modulator driver is directly connected to the spectrum analyzer. Finally, $V_\pi = (\pi/\alpha(L))V_{dev}$ is calculated. The $V_\pi L$ product is calculated by multiplying the $V_\pi$ value with the aperture length ($L$) of the devices.